# The Unified Transition Stages in Linearly Stable Shear Flows

**Jianjun Tao\* and Xiangming Xiong**

*e-mail address of presenting authors: **jjtao@pku.edu.cn**

**Abstract** Abrupt transition to turbulence may occur in pipe and channel flows at moderate flow rates, an unexpected event according to linear stability theory, and has been an open problem in fluid dynamics for more than a century. Extensive numerical simulations and statistical analyses of the plane-Poiseuille flow have been performed. A sequence of transition stages, which includes the equilibrium localized turbulence, the temporally persistent turbulence and the uniform turbulence, is identified. The spread of turbulent band is mainly caused by an oblique extension at moderate Reynolds numbers and band splitting at high Reynolds numbers. The small scale regime of a turbulent band coincides with the oblique swirl region of the mean flow in planes parallel to the walls. Furthermore, this transition scenario is shown quantitatively to be universal for channel and pipe flows in terms of a locally defined Reynolds number.

## 1. Instructions

Finite-amplitude disturbances may trigger an abrupt transition to turbulence in pipe and channel flows at moderate flow rates, for which the base flows are expected to be stable according to linear stability theory. The transition route has been a long-standing open problem since the experimental work of Reynolds (1883). Recent experiments and numerical simulations of Hagen-Poiseuille (pipe) flow (HPF) and of plane-Couette flow (PCF) have shown some similarities between their transition processes, suggesting that they may follow the same route to turbulence (Manneville 2008, Eckhardt 2008, Eckhardt 2011). However, a direct quantitative validation of the universality of the transition scenario in different shear flows is still difficult mainly because of two obstacles. The first one is the definition of the control parameter. Traditionally, the flow state is described by the Reynolds number Re=$UL/v$, where $U$ and $L$ are the characteristic velocity and length scales, and $v$ is the kinematic viscosity of the fluid. For different flows the Reynolds number is defined with different time and spatial scales and hence a benchmark for comparison doesn't exist. A transition from laminar to turbulent flow is often observed near Re of about 2000 for pipe flow and 340 for plane-Couette flow. The second obstacle lies in the possibility that the transition process is composed of several stages and some of them may have not been identified in previous studies. In comparison with the systematic investigations of HPF (for reviews, we refer to Eckhardt, Schneider, Hof, Westerweel 2007, Mullin 2011) and of PCF (Manneville 2008, Tuckerman, Barkley 2011, Manneville 2012), the corresponding study on transition stage of plane-Poiseuille flow (PPF), a pressure-driven shear flow between two parallel plates, is still rudimental (Carlson, Widnall, Peeters 1982). In order to simplify the spatial-temporal complexity of PPF and to reduce the computing costs, a slender computational domain was used in a recent study (Tuckerman 2012), and one of its periodic directions is chosen to be of the minimal size necessary to sustain turbulence. Considering that the perturbations in plane-Poiseuille flow evolve both in the streamwise and the spanwise directions, the first aim of this study is to use a large computational domain to identify the sequence of transition stages of PPF, and the second one is to validate the universality of the transition scenario in linearly stable shear flows.

## 2. Results

The incompressible Navier-Stokes equations are solved by a spectral code (Chevalier, Schlatter, Lundbladh, Henningson 2007), where the velocity field is expanded in a basis of Fourier modes (in the streamwise $x$- and spanwise $z$-directions) and Chebyshev polynomials (in the wall-normal or transverse direction $y$). The boundary conditions are periodicity in $x$ and $z$ and no-slip at the walls ($y = \pm h$). The base-flow solution is $U_0(y) = U_M(1 - y^2/h^2)$, and the Reynolds number is defined as Re = $U_M h/v$, where $v$ is the kinematic viscosity and $U_M$ is the maximum velocity of base flow. The flow parameters are nondimensionalized with the characteristic velocity $U_M$ and length scale $h$, and the dimensionless flow rate is constant.

### 2.1. Temporally persistent turbulence

The most important issue for the abrupt transition may be the determination of the critical Reynolds number for temporally persistent turbulence. If perturbations are periodic or diminish at boundaries of a flow domain, the volume-averaged growth rate of the disturbance kinetic energy only depends on the energy transferred from the parallel basic flow $U_0(y)$ and the volume-averaged kinetic energy dissipation rate $\varepsilon$ (Serrin 1959):

$$\frac{1}{V}\frac{d}{dt}\int_V kdV = -\frac{1}{V}\int_V (uv\frac{dU_0}{dy})dV - \varepsilon + \frac{4}{3\mathrm{Re}} \quad (1)$$
$$= \frac{1}{V}\int_V (U_0\boldsymbol{u}\cdot\nabla u)dV - \varepsilon + \frac{4}{3\mathrm{Re}}$$

where $V$ is the volume of the field, $\boldsymbol{u}(u,v,w)$ is the disturbing velocity, and $k=\frac{1}{2}\boldsymbol{u}\cdot\boldsymbol{u}$ is the kinetic energy of the disturbance. It is noted that $\varepsilon$ is never less than $4/(3\mathrm{Re})$, the laminar kinetic-energy dissipation rate. According to the Reynolds averaged Navier-Stokes equation, the time-averaged inertia term is $\overline{\boldsymbol{u}\cdot\nabla\boldsymbol{u}} = -\nabla\overline{p} + \nabla^2\overline{U}/\mathrm{Re}$ for a statistically steady parallel flow, where $\overline{U}=(\overline{U}(y),0,0)$ and $\nabla\overline{p}$ are the mean velocity profile and the mean pressure gradient, respectively. During the initial period of the transition, $\overline{U}$ and $\nabla\overline{p}$ only deviate slightly from their basic-flow values and are weak functions of Re, hence the time-averaged value of the integral on the right side of equation (1), the kinetic energy input rate, can be approximated as $A+B/\mathrm{Re}$, where $A$ and $B$ are constants. If the perturbations are temporally persistent and a statistically steady or equilibrium state is reached, the time-averaged growth rate should be zero, and equation (1) becomes $\overline{\varepsilon}\,\mathrm{Re} = A\,\mathrm{Re} + (B+4/3)$, a linear relation between the time-averaged value $\overline{\varepsilon}\,\mathrm{Re}$ and Re.

In order to verify this linear relation, a statistically steady and uniform turbulent state was obtained first at Re=2000 in a large computational domain (320 $h$, 2$h$, 240 $h$) with spectral mode numbers ($nx$, $ny$, $nz$)= (768, 65, 1024). Then the Reynolds number is decreased gradually to look for turbulent states that can persist for at least 7000 time units. It is shown in Figure1(a) that at large Reynolds numbers the relation between Re $\overline{\varepsilon}$ and Re is well fitted by a straight dark line:

$\overline{\varepsilon}\,\mathrm{Re} = 0.00177\,\mathrm{Re} - 0.188$ or $\overline{\varepsilon} = 0.00177 - 0.188/\mathrm{Re}$. (2)

Accordingly, the integral term on the RHS of equation (1) can be estimated as $A+B/\mathrm{Re}$ =0.00177-1.521/Re. Based on the inset of Figure1(a), $\overline{\varepsilon}$ will be consistent with the fitted dark line until $Re$ reaches 1000±50 from above, which corresponds to the threshold of temporally persistent turbulence, and agrees very well with the experimental observation (Carlson 1982). For Hagen-Poiseuille (pipe) flow and plane-Couette flow, this threshold was determined as 2040 ± 10 (Avila, Moxey, de Lozar, Avila, Barkley, Hof 2011) and 325 (Shi, Avila, Hof 2013), respectively, by another method studying the mean lifetime of localized turbulence before decaying and splitting. In addition, sustained turbulent bands in PCF were observed in experiments at Re=340 (Prigent, Grégoire, Chaté, *et al*. 2002, Prigent, Grégoire, Chaté, et al. 2003), 325±5 (Dauchot, Daviaud 1995) and in large-domain simulations at Re=330 (Duguet, Schlatter, Henningson 2010).

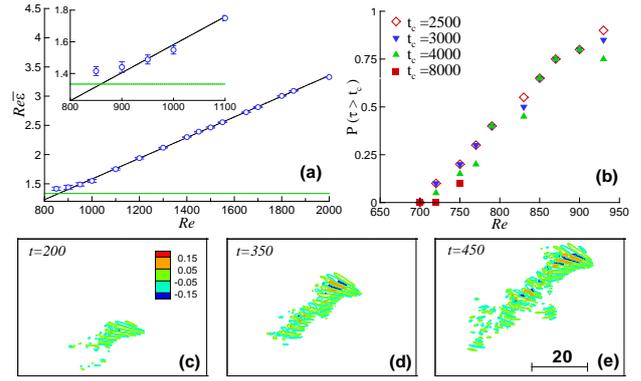

Figure 1 (a) Volume-averaged mean kinetic energy dissipation rate $\overline{\varepsilon}$ as a function of Re. The data at moderate Reynolds numbers are shown in the inset. The temporal variation ranges are indicated by error bars, and the green horizontal line represents the value 4/3 of the laminar state. (b) Survival probability $P$ (lifetime τ> $t_c$) of localized turbulence at low Re. τ is the computation time before the volume-averaged kinetic energy $k$ is less than $10^{-4}$. The temporal evolution of a turbulent band triggered by a local disturbance at Re=900 is shown by iso-contours of the transverse velocity $v$ in (c)-(e).

Oblique turbulent stripes or bands separated by laminar regimes were observed in previous experiments (Carlson, Widnall, Peeters 1982) and numerical simulations (Tsukahara, Seki, Kawamura, Tochio 2005). These patterns may be triggered at moderate Reynolds numbers by the transient growth mechanism (Schmid, Henningson 2001), but cannot survive for a long time. 20 transient states of a field with turbulent bands obtained at Re=1000 were used as initial fields to check their survival probability $P$ (lifetime τ> $t_c$) at low Reynolds numbers in a computational domain of (120$h$, 2$h$, 100$h$) with spectral mode numbers of ($nx$, $ny$, $nz$)= (512, 65, 768). It is shown in Figure 1(b) that $P$ statistically increases with Re as Re>720 and decreases with τ, indicating a transient property of such localized turbulence. $P$ decreases slow for large τ, e.g. two cases at Re=750 can survive at least 8000 time units and $P$ remains as 10% for 4700<τ<8000.

*2.2. Equilibrium localized turbulence*

The statistical properties of equilibrium localized turbulence, such as characteristic length scales and advection velocity, are independent of the initial conditions. When the kinetic energy input rate at equilibrium states 0.00177-1.521/Re is set to be zero, as indicated by the intersection between the dark line and the green line for laminar state in Figure 1(a), one obtains Re=859.3, which is the lower bound for the equilibrium localized turbulence. It was shown in experiments of PPF (Carlson, Widnall, Peeters 1982) that a spot could be triggered by disturbance and developed to two turbulent bands at Re=1000. But for Re<840, only semideveloped transient spot could be formed. It should be noted that the equilibrium localized turbulence near the lower threshold cannot remain a statistically steady state forever. The reason is that a zero energy-input rate corresponds to infinitesimal disturbances, hence the sustained disturbances represent linearly unstable solutions, which contradict the fact that the basic flow is linearly stable at moderate Reynolds numbers. Instead, the perturbed flow may vary stochastically on large-time scales and return to its laminar state eventually as Re is smaller than the threshold for temporally persistent turbulence. Such unsteady states were also observed in simulations of plane-Couette flow for 330>Re≳325 ( Duguet, Schlatter, Henningson 2010). For Hagen-Poseuille flow, the independence of lifetime statistics from initial conditions for transient puffs was found for Re≥1720 (Avila, Willis, Hof 2010).

It is shown in Figure 1(c)-(e) that after an initial evolution period, the locally introduced disturbance at Re=900 evolves to a coherent structure, which extends obliquely while it keeps its spanwise and streamwise length scales at statistically finite values. Consequently, a turbulent band is formed. This phenomenon has also been observed in plane-Couette flows (Duguet, Schlatter, Henningson 2010) for Re≥325, where the bands extend in a zigzag way because its basic flow has no single preferred direction as PPF but opposite directions relative to the middle plane. It is believed that the oblique extension is the main drive to spread the perturbations in channel flows at moderate Reynolds numbers.

In plane Poiseuille flow, an isolated equilibrium turbulent band moves downstream with a statistically constant velocity $u_c<U_M$. As shown in Figure 2 (a), there exist long streamwise vortices in the near-wall region and in the neighborhood of the layer $y=y_c$, where $U_0(y_c)=u_c$. Low-speed streamwise streaks and high-speed streaks are mainly observed for $U_0>u_c$ and $U_0<u_c$, respectively. The main feature of the band is its small scale or turbulent regime, whose dominant structures near the middle plane are quasi-transverse vortices, as shown by the red vortex structures in Figure 2(a). The fluid exchange between the upper and the lower half domains mainly occurs in the turbulent regime (confined by a solid yellow line and a dark line in Figure 2b), and is shown by the iso-contours of the transverse velocity $v$.

A mean flow field was obtained by averaging the transient states in a reference frame moving with $u_c$. It is shown in Figure 2(c) by the arrows that in the midplane the mean disturbance velocity changes amplitude and direction in a tilted band space, an oblique swirl region. Importantly, it is examined that the swirl region of the mean flow coincides with the turbulent regime not only in the midplane but also in any $x$-$z$ plane, indicating the key role of this region in the production of the small-scale structures. Furthermore, the mean flow around the swirl region brings disturbances into the unperturbed neighborhood in the tilted direction, and then extends the localized perturbations into a longer turbulent band.

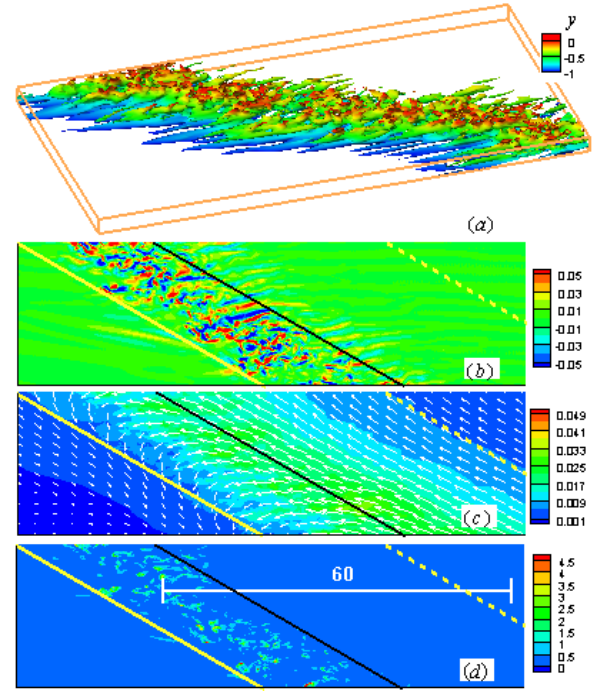

Figure 2 Three-dimensional structures in the lower half space (a) and two-dimensional properties in the middle plane (b)-(d) of an 'equilibrium' localized turbulence at Re=950. (a) Iso-surface of the amplitude of disturbance vorticity at 0.67 with color contours scaled to the vertical coordinate $y$. (b) Iso-contours of the disturbing transverse velocity. (c) Iso-contours of the disturbing kinetic energy averaged over 50 transient states with a time step of 50 in a moving reference frame at speed $0.75U_M$. The in-plane mean flow is shown by arrows. (d) Iso-contours of the kinetic-energy dissipation rate multiplied by $Re$. The tilted solid and dashed yellow lines indicate the boundaries where the disturbance kinetic energy is about 0.005, and the tilted solid dark line represents the ridge of the mean disturbance kinetic energy as shown in (c).

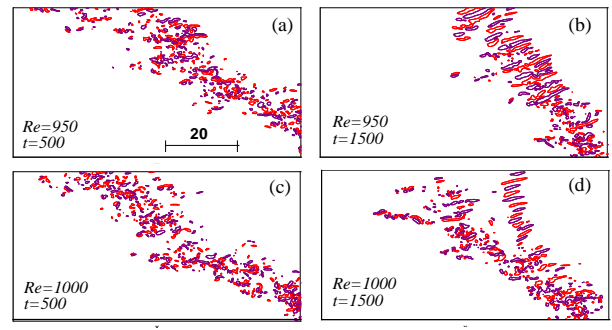

Figure 3 Temporal evolutions of turbulent bands at different Reynolds numbers with the same initial flow field. The iso-contours of the transverse velocity are shown at ±0.05.

At high Reynolds numbers the turbulent band can be observed to split into branches. It is shown in Figure 3

for the same initial field that after a 1500-time-unit computation band splitting occurs at Re=1000, while it still remains an isolated band at Re=950. The band splitting is much more efficient than the oblique extension in spreading perturbations, and thus sustains statistically the localized turbulent structures.

It should be noted that, when the localized turbulent bands occupy the whole computational domain by oblique extension and splitting, an unphysical factor, the influence of the periodic boundaries in the streamwise and spanwise directions can no longer be neglected. Especially, when the lengths of the computational domain are nearly integral times of the characteristic length scales of the turbulent band, well-arranged oblique bands may be observed, and the correlation between bands becomes strong. Based on autocorrelation analysis of the spanwise disturbance velocity $w$, the statistically characteristic streamwise and spanwise spaces between bands can be determined. For example, simulations show that the streamwise spacing varies around 63, 48 and 44 at Re=1300, 1500 and 1700, respectively. Accordingly, the dominant oblique angle of bands relative to the streamwise direction can be calculated and is found to change from $\pm 36°$ at Re=1300 to $\pm 26°$ at Re=1700. When $Re$ is increased further, the absolute value of oblique angle increases slightly and is difficult to identify because the large-scale pattern has become spatially intermittent.

### 2.3. Uniform turbulence

For pressure-driven flows, the perturbations on the axis of HPF or the middle plane of PPF are more uniform than at other places because the mean shear is zero there. In addition, the transverse velocity $v$ reflects more features of small-scale structures than large-scale ones because its amplitude and characteristic transverse length scale are constrained by the side wall(s). A continuous transition from the intermittent turbulent-laminar state to uniform turbulence was identified at Re $\approx$ 2600 for Hagen-Poiseuille flow by analyzing the probability density function (pdf) of a single-point transverse velocity on the pipe axis (Moxey, Barkley 2010), and a similar transition was found numerically in plane Couette flows (Tuckerman, Barkley 2011, Tuckerman, Barkley, Moxey, et al 2009). In the present study, we have calculated the pdf of the normalized transverse velocity $v/\sigma$ in the entire midplane, where $\sigma$ is the standard error of $v$. It is well known that the pdf is Gaussian for developed turbulence, and it is shown in Figure 4(a) that the pdf(0) decreases with $Re$, and is very close to the Gaussian value 0.4 as $Re \geq 1400$. Because of the strong shear in the near-wall region, perturbations are still spatially intermittent at Re=1300 (Figure 4c), but become uniform in the $x$-$z$ planes at $Re$=1800 (Figure 4d). In order to further rule out the effect of periodic boundaries, we examine the behavior of small structures represented by the kinetic-energy dissipation rate as shown in Figure 2(d). The area averaged kinetic energy dissipation rate $\varepsilon_0$ is defined as $\varepsilon_0 = \frac{1}{S}\int_S \|\nabla \boldsymbol{u}\|^2 dS$, where $S$ is the area of the midplane. It is shown in Figure 4(b) that $\varepsilon_0$ increases with $Re$ for spatially intermittent flows before reaching its maximum value, then decreases with $Re$ in almost the same manner as that of flows with uniform perturbations even in the near-wall regions, e.g. Re=1800, 1850 and 2000. Considering the features shown in Figure 4(a) and (b), the threshold of uniform turbulence is defined as $1350 \pm 50$, where $\varepsilon_0$ reaches its maximum value. In experiments of PPF, a continuously turbulent signal was obtain in the $Re$ region of 1350 (Patel, Head 1969).

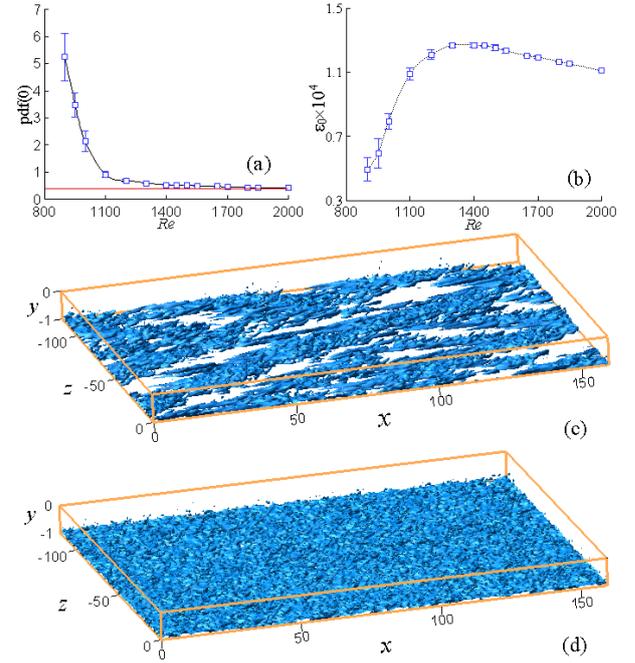

Figure 4 (a) pdf(0) of the normalized transverse velocity $v/\sigma$ in the midplane, where $\sigma$ is the standard error of $v$. The red line represents the corresponding Gaussian value. (b) Kinetic-energy dissipation rate averaged in the middle plane $\varepsilon_0$ as a function of Re. Temporal variation ranges are shown by the error bars. The iso-surface of the vorticity $|\omega|=9$ in one eighth of the computational domain at Re=1300 and 1800 are shown in (c) and (d), respectively.

### 2.4. Local Reynolds number

As discussed above, shear flows with different boundary geometries (pipe or channel) and driving forces (shear-driven or pressure-driven) show similar transition stages. A simple explanation of this similarity is the fact that the dynamics governing the movement of fluid element is the same. Investigators have tried for more than half a century to look for a universal control parameter based on governing equations (Ryan, Johnson 1959, Hanks 1963, Tao, Chen, Su 2013). Based on the energy equation and intuitive arguments, Ryan and Johnson (1959) proposed a local control parameter $Z_{max} = \sqrt{\frac{4}{27} \frac{R U_M}{\nu}}$, representing a ratio of input energy to energy dissipation for a fluid element, to define the transition threshold of non-Newtonian fluids in pipe flows. Hanks (1963) extended this concept to plane-Poiseuille flow by suggesting another locally defined control parameter $\kappa = \frac{1}{2}\rho \left| \nabla(U_0^2)/\nabla p_0 \right|_{max}$, where the subscript 'max' represents the maximum value in the cross section. $\kappa$ is equivalent to $Z_{max}$ for pipe flow ($Z_{max} = 2\kappa$) but cannot be used for shear-driven flows, e.g. plane-Couette flow, where $\nabla p_0 = 0$. Inspired by these pioneering works, a local Reynolds number $Re_L$ was proposed recently (Tao, Chen, Su 2013) based on the energy equation of a fluid element:

$$Re_L = \left| \rho R_h U_0 \frac{dU_0}{dx_2} / \tau_{w,0} \right|_{max} \quad (3)$$

where $x_2$ is the transverse coordinate, $\tau_{w,0}$ is the viscous shear stress of the base flow on the wall, and the hydraulic radius $R_h$ is a ratio of the cross-section area to the wetted perimeter. $U_0 dU_0/dx_2$ has a dimension of force per unit mass. According to equation (1), the energy input rate from the basic flow to the perturbations is directly related to $U_0(x_2)$ and $dU_0/dx_2$, hence a fluid element with larger $|U_0 dU_0/dx_2|$ has a stronger ability to absorb kinetic energy from the basic flow. In addition, fluid elements may have the same disturbance kinetic-energy dissipation rate when they are perturbed by the same viscous force of $\tau_{w,0}/(\rho R_h)$. Therefore, $Re_L$ represents the maximum ratio of the energy input to the energy dissipation for fluid elements in the entire cross section. For detailed discussions about $Re_L$ we refer to (Tao, Chen, Su 2013). It is easy to verify that the relations between $Re_L$ and the traditional Reynolds number Re for the plane-Poiseuille flow (PPF), pipe or Hagen-Poiseuille flow (HPF) and plane-Couette flow (PCF) are $Re_L = \frac{2}{3\sqrt{3}} \frac{U_M h}{\nu} = \frac{2}{3\sqrt{3}} Re$, $Re_L = \frac{1}{3\sqrt{3}} \frac{U_M R}{\nu} = \frac{1}{3\sqrt{3}} Re$ and $Re_L = \frac{U_M h}{\nu} = Re$ respectively, where $R$ is the radius of the pipe. Therefore, the local Reynolds number $Re_L$ has the same expressions as $\kappa$ for pressure-driven pipe and channel flows, but $Re_L$ is also applicable for shear-driven flows, such as plane-Couette flow.

The related threshold values discussed above for PPF, PCF and HPF are summarized in Table I. Since it was shown numerically (Tuckerman, Barkley 2011) that different tilt angles of the narrow computational domains relative to the streamwise direction lead to different transition scenarios for PCF, only experiments and simulations with large computational domains are included in Table I. The thresholds defined with traditional Reynolds number are completely different for the various flows with a relative difference about 420% - 540%. When the local Reynolds number $Re_L$ is applied, the two Poiseuille flows (HPF and PPF) have almost the same threshold values for all three transition stages with a percentage difference less than 8%. For the three kinds of flows, the maximum percentage difference between the thresholds is still less than 27%. These surprising consistencies confirm the universality of the transition scenario of these linearly stable shear flows.

Table I. The threshold Reynolds numbers of the abrupt transition in different shear flows.

| Flow type | Equilibrium localized turbulence | | |
|---|---|---|---|
| PCF | 325[6] | N | 325 |
| HPF | 1720[2] | N | 331 |
| PPF | ≥840[3] | E | ≥323.3 |
| | 859.3(present) | N | 330.7 |
| | Temporally persistent turbulence | | |
| PCF | 340[16,17] | E | 340 |
| | 325±5[5] | E | 325±5 |
| | 330[6] | N | 330 |
| HPF | 2040±10[1] | N,E | 392.5±1.9 |
| PPF | ≃1000[3] | E | ≃385 |
| | 1000±50(present) | N | 385±19.2 |
| | Uniform turbulence | | |
| HPF | 2600[13] | N | 500.4 |
| PPF | ≃1350[15] | E | ≃519.6 |
| | 1350±50(present) | N | 519.6±19.2 |

Note: The middle column lists the values defined with the traditional Reynolds number and the corresponding local Reynolds numbers $Re_L$ are shown in the right column. E and N represent experimental and numerical results, respectively.

**Conclusion**

Localized turbulence is a key feature at the initial stages of the abrupt transition, and exhibits distinct characteristic features for different flow geometries. For example, at moderate Reynolds numbers turbulent bands in PPF may extend obliquely in the $x$-$z$ plane while puffs in HPF are constrained by the circular wall. The turbulent region near the midplane of PPF shows plenty of transverse momentum exchange and quasi-transverse vortex structures, which are different from the streamwise structures in the near-wall region. Consequently, PPF cannot be understood simply as a composition of two PCF side by side. On the other hand, the dynamic behavior of each fluid element is controlled by the same constitutive relation. When a control parameter $Re_L$ defined with local flow properties is applied, the thresholds of different transition stages between these flows show quantitative consistency, indicating that the critical state of every stage is determined, to some degree, by the fluid elements which are most vulnerable to external perturbations in a cross section. In addition, the approach here to determine the onset of sustained turbulence is based on the balance between the total energy input rate and the kinetic energy dissipation rate, and hence is expected to be equally applicable for other viscous shear flows.

**Acknowledgments:** We wish to thank many cited authors for insightful discussions. The simulation code SIMSON from KTH and the help from P. Schlatter, L. Brandt and D. Henningson are gratefully acknowledged. This work has been supported by the National Natural Science Foundation of China (11225209, 10921202).

**References**:

[1] Avila, K., D. Moxey, A. de Lozar, M. Avila, D. Barkley, B. Hof (2011) "The onset of turbulence in pipe flow". *Science* 333, pp. 192–196.

[2] Avila, M., A.P. Willis, B. Hof (2010) "On the transient nature of localized pipe flow turbulence". *J. Fluid Mech.* 646, pp. 127–136.

[3] Carlson, D.R., S.E. Widnall, M.F. Peeters (1982) "A flow-visualization study of transition in plane Poiseuille flow". *J. Fluid Mech.* 121, pp. 487–505.

[4] Chevalier, M., P. Schlatter, A. Lundbladh, D.S. Henningson (2007) "SIMSON a pseudo-spectral solver for incompressible boundary layer flows", Technical Report, KTH, Stockholm.

[5] Dauchot, O. and F. Daviaud (1995) "Finite amplitude perturbation and spots growth mechanism in plane Couette flow". *Phys. Fluids* 7, 335–343.

[6] Duguet, Y., P. Schlatter, D. Henningson (2010) "Formation of turbulent patterns near the onset of transition in plane Couette flow". *J. Fluid Mech.* 650, pp. 119-129.

[7] Eckhardt B. (2008) "Turbulence transition in pipe flow: Some open questions". *Nonlinearity,* 21, pp. T1–T11.

[8] Eckhardt B. (2011) "A critical point for turbulence" *Science,* 333, pp. 165–166.

[9] Eckhardt, B., T.M. Schneider, B. Hof, J. Westerweel (2007), "Turbulence Transition in pipe flow". *Annu. Rev. Fluid Mech.* 39, pp. 447-468.

[10] Hanks, R.W. (1963) "The laminar-turbulent transition for flow in pipe, concentric annuli, and parallel plates". *A.I.Ch. E. J.* 9, pp. 45–48.

[11] Manneville, P. (2008) "Understanding the sub-critical transition to turbulence in wall flows". *PRAMANA. J. Phys.* 70, pp.1009–1021.

[12] Manneville, P.(2012) "On the growth of laminar - turbulent patterns in plane Couette flow". *Fluid Dyn. Res.*44, 031412.

[13] Moxey, D. and D. Barkley (2010) "Distinct large-scale turbulent-laminar states in transitional pipe flow". *PNAS* 107, pp. 8091-8096.

[14] Mullin, T. (2011) "Experimental studies of transition to turbulence in a pipe". *Annu. Rev. Fluid Mech.* 43, pp. 1-24.

[15] Patel, V.C. and M.R. Head (1969) "Some observations on skin friction and velocity profiles in fully developed pipe and channel flows" *J. Fluid Mech.* 38, pp. 181–201.

[16] Prigent, A., G. Grégoire, H. Chaté, *et al.*(2002) "Large-scale finite-wavelength modulation within turbulent shear flows". *Phys. Rev. Lett.* 89, 014501

[17] Prigent, A., G. Grégoire, H. Chaté, et al. (2003) "Long-wavelength modulation of turbulent shear flows". *Phys. D*, 174, pp. 100–113.

[18] Reynolds O. (1883) *Proc. R. Soc. London* 35, 84.

[19] Ryan, N.W. and M. Johnson (1959) "Transition from laminar to turbulent flow in pipes". *A.I.Ch. E. J.* 5, pp. 433–435.

[20] Schmid, P.J., D.S. Henningson, *Stability and Transition in shear flows.* Springer, 2001

[21] Serrin, J. (1959) "On the stability of viscous fluid motions". *Arch. Rational Mech. Anal.* 3, pp. 1-13 .

[22] Shi, L., M. Avila, B. Hof (2013) "Scale invariance at the onset of turbulence in Couette flow". *Phys. Rev. Lett.* 110, 204502.

[23] Tao, J.J., S.Y. Chen, W.D. Su (2013) "Local Reynolds number and thresholds of transition in shear flows". *Sci. China-Phys. Mech. Astron.* 56, pp. 263-269.

[24] Tsukahara, T., Y. Seki, H. Kawamura, D. Tochio (2005) "DNS of turbulent channel flow at very low Reynolds numbers" *Proc. 4th Int. Symp. on Turbulence and Shear Flow Phenomena.* 935–940.

[25] Tuckerman, L. (2012) "Turbulent-laminar banded patterns in plane-Poiseuille flow". *Proc. XXIII ICTAM*, 19–24 August, Beijing, China.

[26] Tuckerman, L., D. Barkley (2011) "Patterns and dynamics in transitional plane Couette flow" *Phys. Fluids* 23, 041301.

[27] Tuckerman, L.S., D. Barkley, D. Moxey, et al (2009) "Order parameter in laminar-turbulent patterns". In: B. Eckhardt, ed. *Advances in Turbulence* XII. NY: Springer, 132, pp. 89–91.

**Author Information**

*Prof. Jianjun Tao, SKLTCS and CAPT, Department of Mechanics and Engineering Science, College of Engineering, Peking University, Beijing 100871, China*

*Xiangming Xiong, SKLTCS and CAPT, Department of Mechanics and Engineering Science, College of Engineering, Peking University, Beijing 100871, China*